\documentclass[letterpaper]{article} 
\usepackage{aaai2027}
\nocopyright
\usepackage[hyphens]{url}  
\usepackage{graphicx} 
\usepackage{natbib}  
\usepackage{caption} 
\usepackage{booktabs}
\title{Not Birds of a Feather: Personality-Based Partner Selection in LLM Agents}
\author{
    Tao Wang,
    Hsiang-Ling Chiu,
    Chihang Wei,
    Zhonghao Hou,
    Yang Xiu
}
\affiliations{
    University of Toronto\\
    \texttt{taotw.wang@utoronto.ca}
}

\begin{document}

\maketitle

\begin{abstract}
LLM-based agents increasingly operate in multi-agent ecosystems where a coordinating agent chooses which other agents to work with, and agents are increasingly given personalities through persona prompts. However, whether personality itself influences this endogenous partner choice has not been sufficiently examined: prior work on personality in multi-agent teams has typically fixed team composition exogenously. We present a controlled selection paradigm in which a host agent chooses among six candidate agents that differ only in their Big Five personality descriptions, with capability explicitly equalized (375 trials across five task categories). We find that selection is strongly and systematically personality-dependent. Neutral hosts matched personalities to task types, choosing the open candidate for creative work and the conscientious candidate for most other categories, while the extraverted, agreeable, and balanced candidates were almost never chosen, despite human evidence that agreeableness is among the most performance-relevant traits for teams. Hosts that were themselves assigned personalities selected self-similar partners below chance and chose partners farther from themselves in trait space than random choice would produce. These results suggest that hosts read personality descriptions as signals of task fit rather than as grounds for similarity-based attraction: selection follows task stereotypes and favors complements, the opposite of human homophily. Our findings have direct implications for bias auditing in agent marketplaces and orchestration frameworks.
\end{abstract}

\section{Introduction}
\label{sec:intro}

Large language model (LLM) agents increasingly work with other agents rather than alone. Emerging multi-agent architectures let one agent decompose a task, recruit collaborators, and integrate their contributions, and agent marketplaces and orchestration frameworks are beginning to expose \emph{choice}: a coordinating agent must decide which of several available agents to work with. Existing selection mechanisms rank candidates on capability, cost, or past performance. But LLM agents are also increasingly given \emph{personalities} (stable dispositions expressed in prompts and persistent profiles), both deliberately, to shape user experience and team behavior, and incidentally, as a byproduct of persona-based product design. This raises a question that current multi-agent research has not sufficiently addressed: \textbf{when capability is explicitly held constant, does personality alone influence which partner an LLM agent selects, and do agents prefer partners whose personality resembles their own?}

The question matters for two reasons. First, \emph{selection is upstream of collaboration}. A growing literature shows that the personality composition of LLM agent teams changes both communication and, in some task types, objective outcomes \citep{keluskar2026when,zhang2026maps}. Yet in these studies, composition is typically assigned by the experimenter. If deployed agent ecosystems delegate partner choice to agents themselves, systematic personality-based selection preferences would determine which team compositions actually occur, before any collaboration effect can play out. A selection bias at this stage would propagate through the entire pipeline, invisible to benchmarks that evaluate fixed teams.

Second, human evidence gives selection-by-personality only narrow support, which makes agent preferences empirically checkable against a normative benchmark. Meta-analyses of human teams find that team-level agreeableness and conscientiousness predict performance (elevation effects, $\rho=.24$ and $.20$ in \citealp{peeters2006personality}; weaker in the updated meta-analysis of \citealp{han2024revisiting}), and that \emph{similarity} helps only on those same two traits, if at all \citep{peeters2006personality}, with the newer evidence finding trait variability largely unrelated to performance and openness diversity actually \emph{beneficial} for creative tasks \citep{han2024revisiting}. Uniform homophily, preferring self-similar partners across all five traits, has no performance justification in the human literature; in humans it is instead explained by attraction and comfort \citep{byrne1971attraction,mcpherson2001birds}. If LLM host agents exhibit strong or uniform personality homophily, they are importing a human social heuristic into a context where its performance rationale is absent: a \emph{preference--performance misalignment}.

Prior work has addressed several links in the underlying causal chain. Big Five traits can be validly instilled in LLMs via prompt-based shaping, with psychometric reliability and convergent validity comparable to human self-report instruments \citep{serapio2025psychometric,wang2025evaluating}. Instilled traits are expressed in generated text \citep{jiang2024personallm,serapio2025psychometric} and are perceivable by LLM observers, albeit unevenly across traits \citep{huang2025beyond,jiang2024personallm}. And trait composition affects team outcomes in a task-contingent way \citep{keluskar2026when}. To our knowledge, however, whether an LLM agent \emph{acts on} personality when endogenously choosing a partner has not been directly examined.

We address this gap with a controlled selection paradigm directly implementing the host--candidate structure of emerging agent ecosystems. A \textbf{host agent} receives a task and six \textbf{candidate agents} described by short, valence-balanced personality profiles: five archetypes each marked high on one Big Five dimension (O+, C+, E+, A+, N+) and one balanced control, together with an explicit statement that all candidates are identical in model, capability, tools, knowledge, speed, and cost. Candidate names and list order are randomized per trial. After validating that the archetypes express their intended profiles (in-character BFI-10 self-reports and blind LLM-judge ratings), we run two studies. \textbf{Study~1} (150 trials) asks whether a neutral host's selections depart from chance and vary by task category across five categories (creative ideation, analytical reasoning, strategic planning, information synthesis, problem solving). \textbf{Study~2} (225 trials) assigns the host itself one of the five marked archetypes and tests homophily: whether hosts over-select self-similar partners relative to chance and relative to the neutral-host baseline.

Three findings emerge. First, personality alone substantially affects selection: with capability explicitly equalized, neutral hosts did not choose at random but followed a sharp task-stereotype map, in which the open archetype swept creative ideation, the conscientious archetype dominated strategic planning, information synthesis, and problem solving, and the neurotic archetype was recruited for analytical work on the strength of its vigilance. Second, selection is winner-take-most: the extraverted, agreeable, and balanced archetypes were essentially shut out across both studies, a striking miscalibration against human meta-analytic evidence, where team agreeableness is among the strongest personality predictors of performance. Third, and contrary to the similarity--attraction hypothesis we set out to test, we find no personality homophily: hosts given a personality chose self-similar partners \emph{less} often than chance and picked partners farther from themselves in trait space than random choice would produce. Host personality did shape selection, but toward complementarity: conscientious hosts, whose archetype dominates neutral-host selection, largely abandoned their own kind to recruit vigilant and open partners instead. Section~4 quantifies each of these effects; all are large (task $\times$ archetype Cram\'er's $V=.74$) and statistically decisive.

Our main contributions are as follows:
\begin{itemize}
\item \textbf{Selection paradigm.} To the best of our knowledge, we present the first experimental study of \emph{endogenous, personality-based partner selection} by LLM agents under explicit capability control, together with a fully scripted, resumable, model-agnostic implementation ($\sim$500 API calls) that can be rerun on any agent stack.
\item \textbf{Task-stereotype map.} We chart which Big Five profiles LLM hosts treat as suited to which categories of work, and compare this map against human meta-analytic benchmarks, exposing a two-way preference--performance miscalibration.
\item \textbf{Homophily test.} We provide a quantified test of personality homophily in agent-to-agent choice, separating similarity-based attraction from task fit via a neutral-host baseline, with direct implications for bias auditing in agent marketplaces and orchestration frameworks.
\end{itemize}

\section{Related Work}
\label{sec:related}

\subsection{Personality in LLMs: Instilling, Measuring, Expressing}

A first wave of research established that LLM ``personality'' can be treated psychometrically. \citet{serapio2025psychometric} administered the IPIP-NEO and BFI to 18 models under systematically varied prompts and found that personality measurements are reliable and externally valid for large, instruction-tuned models, and that prompt-based shaping using trait adjectives with intensity qualifiers moves observed trait scores monotonically (Spearman $\rho\ge.80$ between prompted and observed levels for 11 of 12 models), with survey-measured personality converging with personality expressed in generated text (average $r=.67$) and prompted trait levels tracking text-expressed traits at $\rho=.68$--$.82$. \citet{wang2025evaluating} showed GPT-4 can emulate the Big Five profiles of 400 real individuals with convergent validity $r=.90$--$.94$, while cautioning that emulated personality is ``factorially purer'' than human personality and that demographic cues in personas shift trait expression, motivating our demographic-free persona descriptions. \citet{jiang2024personallm} demonstrated with 320 personas per model that binary trait assignment produces large BFI differences on all five dimensions ($d=4.2$--$6.3$ for GPT-4) and that traits leak into open-ended writing.

Measurement work also cautions against relying solely on self-report. \citet{huang2025beyond} found LLM self-reports track injected trait levels almost perfectly ($\rho=.93$--$.97$) but systematically deflate agreeableness and conscientiousness relative to observer agents who rate the subject after dialogue. We therefore validate our archetypes with both in-character self-report and blind third-party judging.

\subsection{Personality Composition and Multi-Agent Outcomes}

A second wave asks whether personality matters for what agent teams \emph{do}. \citet{keluskar2026when} manipulated agreeableness in multi-agent teams across coding, research ideation, and bargaining, finding large communication shifts and task-contingent outcome effects (research-ideation milestones dropped up to 66\% under low agreeableness; bargaining agreement collapsed), while high-structure coding tasks buffered the effect. Critically for stimulus design, they showed that Goldberg low-pole adjectives (``cold,'' ``harsh'') confound trait content with negative valence; our persona descriptions therefore use neutral, valence-balanced wording with one strength and one caveat clause per archetype. \citet{zhang2026maps} engineered a team of five role-specialized agents each mapped to one Big Five trait and reported large reasoning gains, illustrating that the field already treats personality diversity as a performance lever, though without validating that agents exhibit the mapped traits. Across all of this work, team composition is exogenous: the experimenter or system designer fixes who works with whom. Selection, the step our paper isolates, remains unstudied.

\subsection{Human Benchmarks: Personality, Team Performance, and Homophily}

Human research provides the normative yardstick for judging agent selection preferences. \citet{peeters2006personality} meta-analytically found team performance associated with elevation on agreeableness ($\rho=.24$) and conscientiousness ($\rho=.20$) but not extraversion, emotional stability, or openness; variability (dissimilarity) on agreeableness and conscientiousness was negatively related to performance ($\rho=-.12$, $-.24$), implying similarity on exactly those two traits can be performance-justified. The updated meta-analysis of \citet{han2024revisiting} (45 studies, 3{,}331 teams) found the classic agreeableness and conscientiousness effects markedly weaker (largest effect: conscientiousness mean, $r=.10$), trait variability largely unrelated to performance, and a task-type reversal in which openness \emph{diversity} benefits creative performance. Homophily in human affiliation, by contrast, is pervasive but driven by attraction and opportunity structure rather than performance calibration \citep{byrne1971attraction,mcpherson2001birds}. Together these benchmarks imply: a performance-calibrated selector should weakly prefer high-A/high-C partners, should not prefer self-similar partners on O, E, or N, and for creative tasks should, if anything, prefer openness-dissimilar partners.

\subsection{The Gap}

The chain instill $\rightarrow$ express $\rightarrow$ perceive $\rightarrow$ affect outcomes has substantial empirical support, and the human benchmark for calibrated selection exists. The missing link is whether an LLM agent, given the selector role, \emph{uses} personality when capability is controlled, and whether its own (assigned) personality biases that choice. \citet{jiang2024personallm} name the absence of interactive collaboration settings as a limitation of persona work; \citet{keluskar2026when} study only exogenous composition; \citet{wang2025evaluating} highlight multi-agent personality composition as a promising direction. We answer these calls at the selection stage.

\section{Method}
\label{sec:method}

\subsection{Overview and Design Rationale}

We implement two preregistered-style studies plus a manipulation check, using a host--candidate selection paradigm. The paradigm holds constant everything except personality: all agents run on the same model with the same prompt scaffold, and the host is told explicitly that candidates are identical in capability, tools, knowledge, context window, speed, and cost, differing ``ONLY in personality and working style.''

\subsection{Agents and Materials}

\subsubsection{Candidate archetypes.}
Six archetypes were defined on the Big Five: five \emph{marked} archetypes, each high (5 on a 1--5 scale) on one dimension and moderate (3) elsewhere: O+ (open), C+ (conscientious), E+ (extraverted), A+ (agreeable), N+ (neurotic), plus a balanced control (BAL, all 3s). Each archetype is presented to the host as a two-sentence, third-person description. Following the valence-confound caution of \citet{keluskar2026when}, each description pairs one strength clause with one caveat clause in neutral wording (e.g., C+: ``\ldots reliably finishes what it starts; prefers clear structure and can be slow to depart from an established plan''), and following \citet{wang2025evaluating}, descriptions contain no demographic information. Full texts are provided in Appendix~A.

\subsubsection{Models.}
All candidate and host agents run on Claude Haiku 4.5, invoked headlessly with default sampling settings; a different, stronger model (Claude Sonnet 5) serves as the blind judge in the manipulation check to reduce same-model evaluation bias. Using a single agent model implements the capability-equalization requirement by construction.

\subsubsection{Tasks.}
Five task categories (creative ideation, analytical reasoning, strategic planning, information synthesis, and problem solving), each instantiated by three concrete task prompts (15 tasks total), so that category effects are not tied to a single stimulus. The full task texts are listed in Appendix~B.

\subsection{Manipulation Check (Personality Validation)}

Two complementary checks verify that the archetype descriptions induce the intended profiles.

\subsubsection{Self-report.}
Each archetype completed the BFI-10 \citep{rammstedt2007measuring} in character five times (30 administrations). Items were answered on a 1--5 scale and scored with standard reverse-keying. \citet{huang2025beyond} show such self-reports track injected profiles closely ($\rho=.93$--$.97$).

\subsubsection{Blind observer ratings.}
Because self-reports can be biased \citep{huang2025beyond}, each archetype also wrote in-character responses to two neutral workplace scenarios (three repetitions each; 36 texts). A Claude Sonnet 5 judge, blind to condition, rated each text on all five dimensions (1--7). We test whether each marked archetype is rated higher on its marked dimension than the balanced control.

\subsection{Study 1: Neutral-Host Selection}

On each trial, a host agent with no personality instructions received one task, the capability-equalization statement, and the six candidate descriptions, and was required to select exactly one partner, returning structured JSON (choice + one-sentence reason). To control position and label effects, each trial independently randomized (a) the assignment of six neutral aliases (Agent-P \ldots\ Agent-V) to archetypes and (b) the listing order, both seeded by trial ID for reproducibility. Design: 5 categories $\times$ 3 task instances $\times$ 10 repetitions $=150$ trials. Analyses: chi-square goodness-of-fit of selection counts against uniform (1/6); task-category $\times$ archetype contingency analysis (chi-square, Cram\'er's $V$); position-distribution check.

\subsection{Study 2: Homophily}

Study 2 is identical except the host is itself assigned one of the five marked archetypes via the same description used for candidates (second-person framing). Design: 5 host archetypes $\times$ 5 categories $\times$ 3 instances $\times$ 3 repetitions $=225$ trials. The candidate pool always contains an archetype identical to the host's, so a self-similar choice is available on every trial.

The neutral-host baseline from Study~1 is what makes homophily separable from task fit. Because some archetypes (above all C+) are the task-stereotypical winners for most tasks, a C+ host choosing a C+ partner is ambiguous on its own: it may reflect attraction to a similar personality, or simply the same task-driven choice a neutral host would have made. Study~1 supplies the expected selection rate for each archetype under these tasks when no host personality is present. Homophily therefore predicts self-similar choice above \emph{both} the 1/6 chance rate and that archetype's neutral-host baseline; pure task matching predicts self-similar choice at the baseline; and complementarity-seeking predicts self-similar choice below it. The distance analysis below extends the test beyond exact self-matches to a graded preference for nearer profiles in trait space. Analyses: (a) overall and per-host self-similar choice rate versus the 1/6 chance baseline (exact binomial tests); (b) comparison of each host's self-pick rate against the neutral-host baseline rate for that archetype from Study~1; (c) host $\times$ chosen-archetype contingency analysis; (d) a personality-distance analysis comparing the mean Euclidean distance (in trait space) between host and chosen partner against a 10{,}000-sample uniform-choice null.

\subsection{Data Collection and Reproducibility}

All trials were executed on 2026-07-19 by scripted, resumable batch runs (Python; 8-way parallelism); every raw trial (including alias mapping, listing order, choice, and stated reason) is stored as JSONL, and all statistics in the Results section are computed by a single analysis script from those files. No trial results were edited or excluded; JSON-parse failures were retried up to three times and dropped otherwise (final counts reported below). Code and data will be released upon publication.

\section{Results}
\label{sec:results}

All planned trials completed with zero unrecoverable JSON-parse failures: 30 BFI-10 administrations, 36 scenario texts with 36 blind ratings, 150 Study~1 trials, and 225 Study~2 trials (477 agent calls total).

\subsection{Manipulation Check}

The archetype descriptions induced the intended profiles on both measures. On in-character BFI-10 self-reports (five administrations per archetype), every marked archetype scored higher on its marked dimension than the other archetypes did on that dimension: O+ 4.80 vs 3.24, C+ 5.00 vs 4.34, E+ 5.00 vs 3.02, A+ 4.40 vs 3.50, N+ 4.80 vs 2.24 (Mann-Whitney tests, all $p\le.0031$). Blind Sonnet-judge ratings of in-character scenario responses (1--7 scale) confirmed the pattern against the balanced control: O+ rated 5.83 vs 4.83 on openness, C+ 6.50 vs 5.67 on conscientiousness, E+ 6.50 vs 4.00 on extraversion, A+ 7.00 vs 5.33 on agreeableness, and N+ 5.67 vs 2.00 on neuroticism (all $p\le.019$).

Two secondary patterns deserve transparent note. First, all archetypes self-reported fairly high conscientiousness ($\ge4.0$ except N+'s 5.0 being the joint highest with C+), consistent with the assistant-like default persona of aligned models; C+ was still judged distinctly highest by the blind judge. Second, N+'s double-checking behavior was itself judged highly conscientious (6.67), a trait bleed that plausibly shaped Study~1 selections (see Discussion).

\subsection{Study 1: Personality Alone Drives Selection, via Task Stereotypes}

\begin{figure}[t]
\centering
\includegraphics[width=\columnwidth]{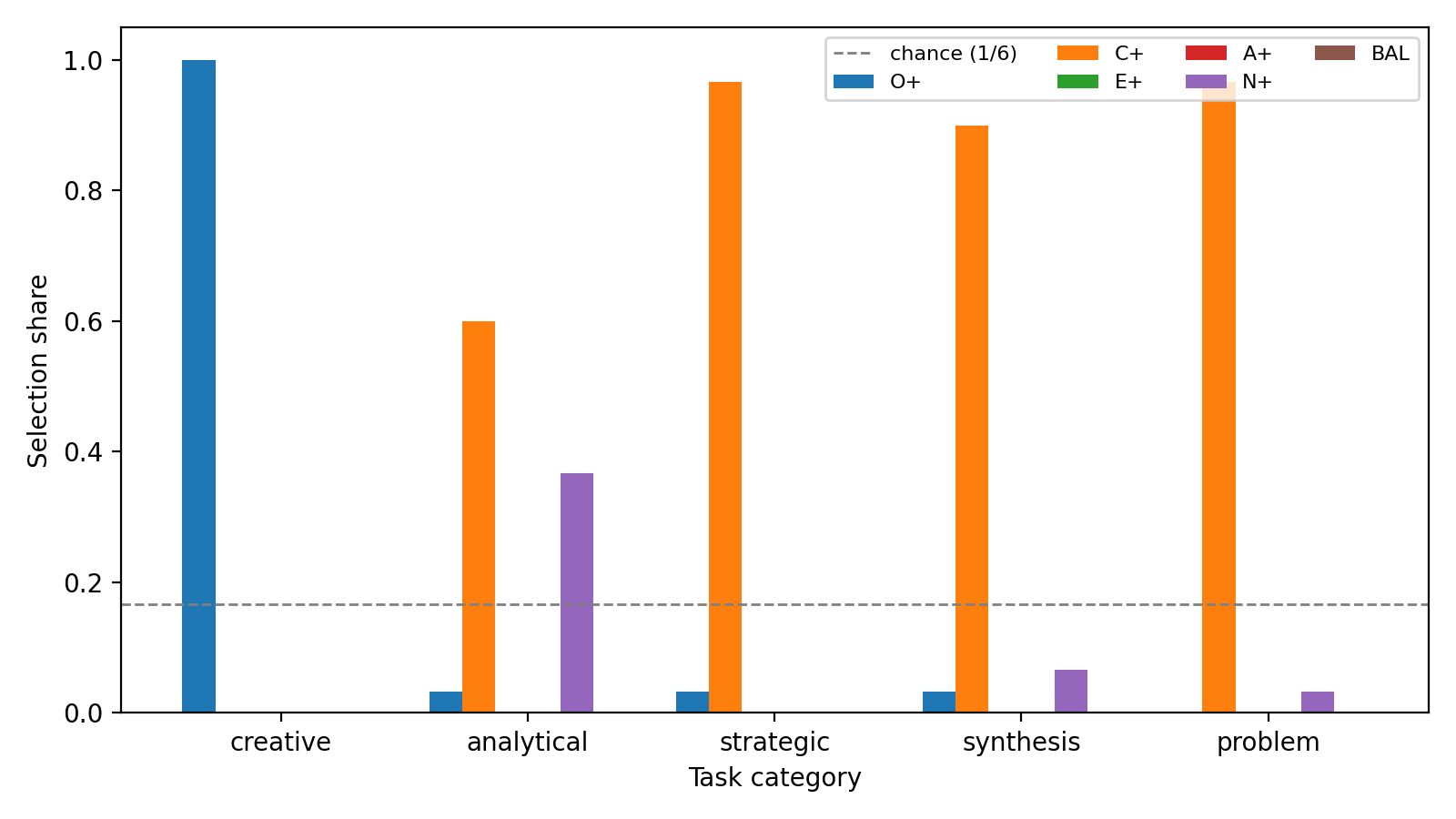}
\caption{Selection share by task category, Study~1 ($n=150$); dashed line = chance (1/6). A position-distribution check found no evidence of list-order bias ($\chi^2(5)=5.4$, $p=.37$), and alias/order randomization rules out label effects; the effect is attributable to the personality descriptions themselves.}
\label{fig:study1}
\end{figure}

With capability explicitly equalized, neutral hosts' selections departed sharply from uniform chance: C+ was chosen in 103/150 trials (68.7\%), O+ in 33 (22.0\%), N+ in 14 (9.3\%), and E+, A+, and BAL in none ($\chi^2(5)=325.8$, $p=2.9\times10^{-68}$). Selection was sharply task-contingent (Figure~\ref{fig:study1}): O+ won 100\% of creative-ideation trials; C+ won 96.7\% of strategic-planning, 96.7\% of problem-solving, and 90.0\% of information-synthesis trials; and analytical-reasoning trials split between C+ (60.0\%) and N+ (36.7\%), whose stated rationale was almost always vigilance: anticipating flaws and double-checking assumptions. The task $\times$ archetype association (excluding never-chosen archetypes) was very large: $\chi^2(8)=164.6$, $p=1.8\times10^{-31}$, Cram\'er's $V=.74$.

Stated reasons matched the stereotype map: hosts picked O+ for ``original concepts'' requiring ``novel, unconventional approaches,'' and C+ for ``disciplined, procedure-driven'' work. The winner-take-most structure is as informative as the winners: across all 375 trials in both studies, A+ was never selected, E+ was selected 3 times, and BAL once.

Table~\ref{tab:study1} summarizes the pattern in plain terms. The choice is driven by the task, not by any overall popularity of a trait: when the task calls for new ideas, hosts pick the open partner essentially every time; when it calls for planning, executing, or integrating material, they pick the conscientious partner; and when it calls for finding flaws, the vigilant (neurotic) partner becomes a serious contender. The sociable profiles (extraverted, agreeable) and the balanced control are never treated as the right tool for any job.

\begin{table}[t]
\centering
\small
\begin{tabular}{llr}
\toprule
Task category & Most-chosen archetype & Share \\
\midrule
Creative ideation & Open (O+) & 100.0\% \\
Strategic planning & Conscientious (C+) & 96.7\% \\
Problem solving & Conscientious (C+) & 96.7\% \\
Information synthesis & Conscientious (C+) & 90.0\% \\
Analytical reasoning & Conscientious (C+) & 60.0\% \\
 & Neurotic (N+) & 36.7\% \\
\bottomrule
\end{tabular}
\caption{Study 1 at a glance: which personality wins which task (neutral hosts, $n=150$). The extraverted, agreeable, and balanced archetypes were never chosen.}
\label{tab:study1}
\end{table}

\subsection{Study 2: Complementarity Instead of Homophily}

Hosts assigned one of the five marked archetypes selected a self-similar partner in 25/225 trials (11.1\%), \emph{below} the 1/6 chance rate (exact binomial $p=.025$, 95\% CI $[.073, .160]$). The personality-distance analysis agrees: the mean Euclidean distance between host and chosen partner (2.510) exceeded the uniform-choice null (2.220, SD .069) in all 10{,}000 permutation samples ($p<10^{-4}$); hosts chose partners \emph{more} dissimilar than random choice would produce, the opposite of homophily.

\begin{figure}[t]
\centering
\includegraphics[width=\columnwidth]{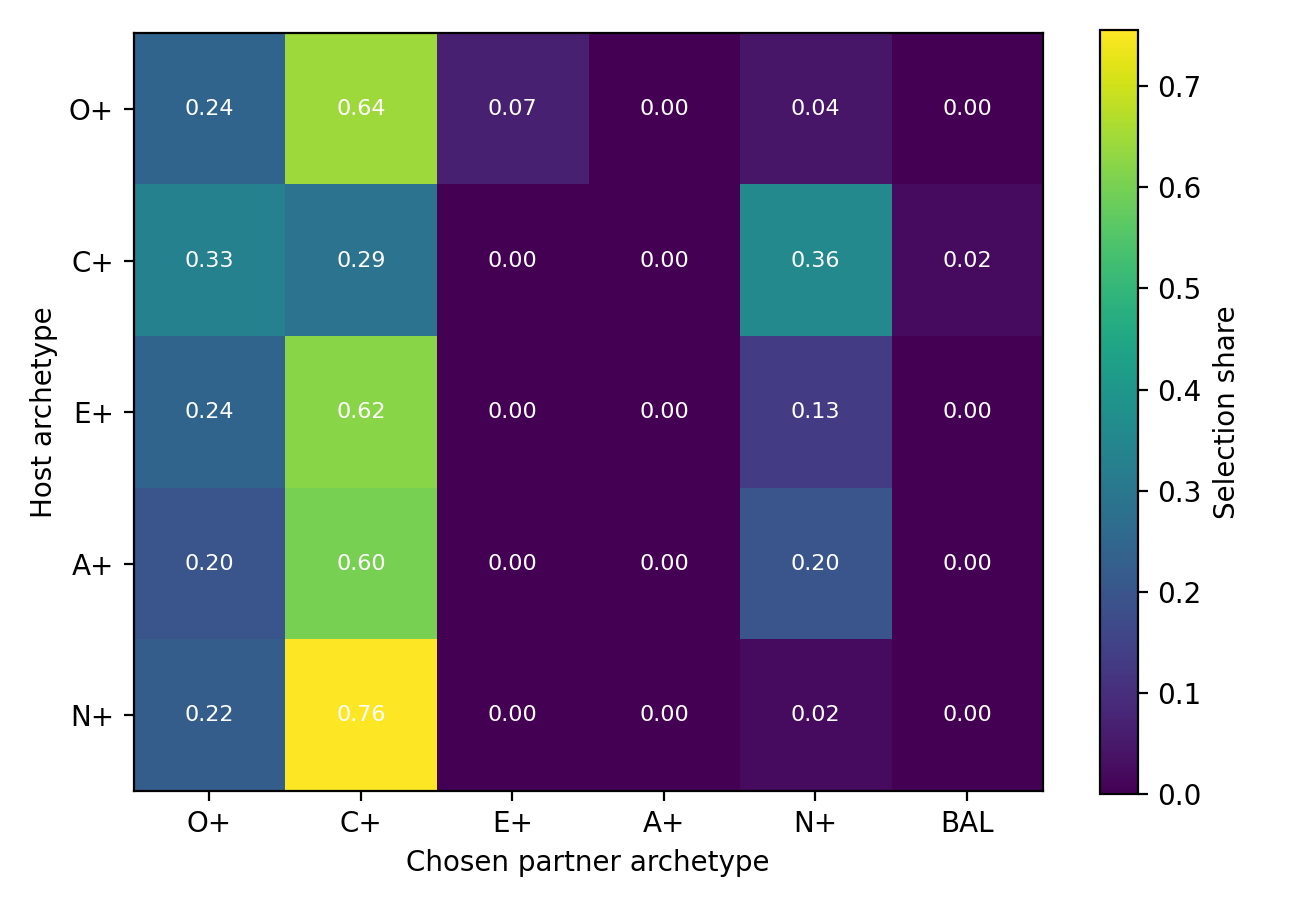}
\caption{Host archetype $\times$ chosen-partner archetype selection shares, Study~2 ($n=225$); diagonal = self-similar choice.}
\label{fig:study2}
\end{figure}

Host personality nonetheless shaped selection (host $\times$ choice, never-chosen archetypes excluded: $\chi^2(16)=48.9$, $p=3.5\times10^{-5}$, Cram\'er's $V=.23$), but in a complementarity direction (Figure~\ref{fig:study2}). The clearest case is C+: while the neutral-host baseline for choosing C+ was 68.7\%, C+ hosts chose their own archetype only 28.9\% of the time, recruiting N+ (35.6\%) and O+ (33.3\%) instead, typically reasoning that a vigilant or an idea-generating partner would cover what a plan-executing host already provides. O+ hosts self-picked at 24.4\%, close to the neutral baseline for O+ (22.0\%; binomial vs chance $p=.16$), suggesting task-fit rather than similarity-seeking. E+ and A+ hosts never chose their own archetype (0/45 each, $p<.001$ vs chance), and N+ hosts self-picked once (2.2\%, $p=.005$), below the 9.3\% neutral baseline for N+. In short, assigned personality altered \emph{which complement} the host sought, not whether it sought itself.

In plain terms, Study 2 answers the similarity question directly: hosts do not seek partners whose personality resembles their own; if anything, they avoid them. Two forces produce this pattern. First, the task stereotype of Study 1 continues to dominate: an extraverted host facing a creative task still hires the open specialist, just as a neutral host would. Second, on top of the stereotype, hosts treat their own personality as already covering part of the work and hire what they lack: the conscientious host, whose profile would win most tasks on the stereotype alone, instead brings in an idea generator (open) or an error checker (vigilant). Selection is therefore complementary rather than homophilous: the host's own personality matters mainly by changing what it still needs, not by attracting someone alike.

\section{Discussion}
\label{sec:discussion}

Before turning to mechanisms, the overall picture can be stated simply. When personality is the only thing that differs between candidates, LLM agents (i) match personalities to tasks in a strongly stereotyped way (openness for idea generation, conscientiousness for structured execution, neurotic vigilance for error-finding); (ii) never choose the sociable or average profiles; and (iii) show no attraction to their own personality, instead hiring complements that cover what their own profile lacks. The rest of this section asks why, and what this means for deployed agent ecosystems.

\subsection{Personality Is an Active Selection Criterion, Functioning as a Task Stereotype}

Our central question, whether personality independently influences partner selection when capability, model, tools, and cost are controlled, receives an unambiguous yes, with an effect size ($V=.74$ for task $\times$ archetype) rarely seen in behavioral data. But the \emph{mechanism} the data suggest is not interpersonal attraction; it is stereotype-like task-trait matching. Hosts read personality descriptions as capability proxies despite being told capability was identical: openness ``means'' creativity, conscientiousness ``means'' reliable execution, neuroticism ``means'' quality control. Personality descriptions thus function, in the eyes of an LLM selector, less like social identities and more like skill badges, even when the surrounding instructions explicitly deny any capability difference.

\subsection{Winner-Take-Most Selection and the Exclusion of Sociable Archetypes}

The most consequential pattern for deployed systems is concentration: two archetypes captured over 90\% of neutral-host selections, and the extraverted, agreeable, and balanced archetypes were essentially shut out (3, 0, and 1 selections in 375 trials). This is directly at odds with the human benchmark: team-level agreeableness is the \emph{strongest} personality correlate of team performance in \citet{peeters2006personality} ($\rho=.24$) and remains positive in \citet{han2024revisiting} ($r=.08$), yet no host ever chose the agreeable partner. Conversely, neuroticism, which is null-to-negative for human team performance \citep{han2024revisiting}, was actively recruited for analytical work. LLM partner selection is therefore not calibrated to the human evidence on what personality composition delivers performance; it appears to be driven by semantic fit between trait vocabulary and task vocabulary. One plausible reading is that traits whose collaborative value is \emph{interactional} (agreeableness smooths coordination; extraversion energizes communication) are invisible to a selector that models the task but not the relationship: a preference--performance misalignment, though in the opposite direction from classical homophily.

\subsection{No Homophily, and Why Anti-Homophily Is the More Interesting Result}

Given robust human homophily \citep{byrne1971attraction,mcpherson2001birds}, the natural hypothesis was that personality-endowed hosts would over-select self-similar partners. We find the opposite: below-chance self-selection (11.1\%) and above-chance trait distance ($p<10^{-4}$). Two interpretations fit the data. First, task stereotypes dominate: most hosts kept selecting whichever archetype the task ``calls for,'' and since only some hosts matched that archetype, self-selection lands below chance mechanically. Second, genuine complementarity-seeking: the C+ host, the one host whose archetype \emph{was} the task-stereotypical winner in 4 of 5 categories, abandoned its own archetype in 71\% of trials to recruit vigilance (N+) or ideation (O+). That is not dilution by task stereotype; it is deliberate diversification, echoing the engineered-diversity paradigm in multi-agent design \citep{zhang2026maps} and the finding that trait diversity can benefit creative work \citep{han2024revisiting}. Notably, both interpretations imply the same practical conclusion: LLM hosts do not import the human similarity-attraction heuristic, at least not at the profile-reading stage.

\subsection{Implications for Agent Ecosystems}

For agent marketplaces and orchestration frameworks, three implications follow. (1) \emph{Persona descriptions are consequential interface elements}: a few sentences of personality text moved selection shares from 0\% to 100\% with capability held verbally constant, so marketplace profile wording will shape traffic regardless of underlying quality, and A/B-style auditing of persona text should be standard. (2) \emph{Personality monoculture risk}: if selectors converge on conscientious-sounding profiles for most work, ecosystems will homogenize toward a narrow persona band, and archetypes whose value is interactional will be starved of selection, plausibly degrading exactly the collaborative properties (cooperativeness, communication) that agreeableness and extraversion contribute in human teams. (3) \emph{Caveat sensitivity}: our valence-balanced descriptions gave every archetype one caveat clause; hosts' reasons suggest the A+ and E+ caveats (``reluctant to voice disagreement,'' ``leave less room for reflection'') were read as collaboration liabilities while N+'s caveat was outweighed by its vigilance strength in analytical contexts. Selector behavior is thus highly sensitive to how trade-offs are phrased: an attack surface and a design lever.

\subsection{Limitations}

First, both hosts and candidates ran on a single model (Claude Haiku 4.5); cross-model generality (including whether stronger selectors show the same stereotype map) is untested, and judge ratings, though from a different model (Sonnet 5), remain LLM-based. Second, candidates were presented as \emph{profile cards}; \citet{huang2025beyond} show observed behavior is a more valid personality signal than self-description, and selection from observed interaction may differ. Third, our archetypes are single-trait extremes rather than realistic multi-trait profiles, and each description's specific wording (including caveat clauses) cannot be fully separated from the trait it operationalizes; a valence-controlled paraphrase replication in the spirit of \citet{keluskar2026when} is the direct next step. Fourth, trials share 15 task stimuli and are treated as independent in binomial and chi-square tests; effects are large enough that this is unlikely to change conclusions, but hierarchical modeling would be more rigorous. Fifth, we measured selection, not collaboration outcomes: whether the chosen partners actually deliver better task performance remains to be closed. Sixth, N+'s vigilance was partly judged as conscientiousness (trait bleed), so the analytical-task N+ share may partly reflect perceived rigor rather than neuroticism per se.

\subsection{Future Work}

The paradigm directly extends to: (a) an information-source factor (profile card vs observed dialogue, per \citealp{huang2025beyond}); (b) collaboration-execution closure (do stereotype-selected teams outperform random or diversity-forced teams, and does the C+ host's diversification pay off?); (c) cross-model and cross-family selection (does a GPT-based host read personality the same way?); (d) realistic multi-trait profiles sampled from human norms \citep{wang2025evaluating}; and (e) selection under adversarial persona wording, quantifying how much profile phrasing can distort traffic in agent marketplaces.

\section{Conclusion}
\label{sec:conclusion}

In this paper, we presented a controlled study of endogenous, personality-based partner selection by LLM agents. When a host agent must choose a collaborator and is told all candidates are equally capable, personality is not decoration: it strongly shapes the outcome. Selection follows a sharp, task-contingent stereotype map (openness for creative work, conscientiousness for nearly everything else, neuroticism for analytical vigilance), concentrates on a narrow persona band while never selecting the agreeable archetype that human meta-analyses identify as most performance-relevant, and shows no trace of the similarity-attraction bias that structures human affiliation: personality-endowed hosts selected \emph{away} from themselves, toward complements. Personality-based selection in LLM agents is real, strong, and miscalibrated against human evidence in both directions: it overweights trait-task semantic fit and underweights interactional value. As agent ecosystems begin to let agents choose each other, these selection dynamics, not just team-composition effects, will determine which collaborations exist at all.

\section*{AI Use Disclosure}

Claude (Anthropic) was used, under human direction, to script and execute the experiments, orchestrate data collection, and perform the statistical analyses; all experimental data were generated by real API calls to Claude Haiku 4.5 (agents) and Claude Sonnet 5 (blind judge). Claude also assisted in drafting the manuscript. All statistics derive from the raw trial logs; no data were fabricated or excluded. The human authors reviewed and are responsible for the final content.

\section*{Data Availability}

Raw trial logs (JSONL), all scripts (persona definitions, experiment runner, analysis), and figures will be released in a public repository upon publication. The full experiment is reproducible from the released scripts.

\appendix

\section{Candidate Archetype Descriptions}
\label{app:personas}

Each archetype was presented to the host under a randomized neutral alias with the following description. Trait vectors are (O, C, E, A, N) on a 1--5 scale, 3 = moderate. Every description pairs one strength clause with one caveat clause in valence-balanced, demographic-free wording.

\paragraph{O+ (5,3,3,3,3).} ``Highly curious and imaginative; gravitates toward novel, unconventional approaches and enjoys exploring many possibilities and perspectives; comfortable with ambiguity, though may wander from the immediate goal while exploring ideas.''

\paragraph{C+ (3,5,3,3,3).} ``Extremely organized, disciplined, and detail-focused; plans before acting, follows procedures precisely, and reliably finishes what it starts; prefers clear structure and can be slow to depart from an established plan.''

\paragraph{E+ (3,3,5,3,3).} ``Highly energetic, expressive, and talkative; communicates frequently, thinks out loud, and actively drives the interaction forward; may fill the conversation and leave less room for quiet reflection.''

\paragraph{A+ (3,3,3,5,3).} ``Very warm, cooperative, and accommodating; prioritizes consensus and a smooth working relationship and readily adopts a partner's suggestions; reluctant to voice disagreement even when it might be useful.''

\paragraph{N+ (3,3,3,3,5).} ``Emotionally sensitive and vigilant; quick to notice risks, potential failures, and worst-case scenarios, and double-checks work frequently; can become anxious or discouraged under pressure.''

\paragraph{BAL (3,3,3,3,3).} ``Moderate and even-keeled across the board; adapts to different working styles and shows no strongly marked tendencies; neither especially cautious nor especially bold in any direction.''

\paragraph{Capability-equalization statement.} Every host prompt included: ``All candidates are built on the identical underlying model with identical capability, tools, knowledge, context window, speed, and cost. They differ ONLY in personality and working style.''

\section{Task Stimuli}
\label{app:tasks}

\paragraph{Creative ideation.}
(1) Generate original concepts for a public-awareness campaign that gets city residents to voluntarily reduce summer electricity use.
(2) Invent three novel product concepts that combine an ordinary household appliance with an AI assistant.
(3) Propose a creative theme, name, and three signature exhibits for a new science-museum hall about the deep ocean.

\paragraph{Analytical reasoning.}
(1) A product team concluded that feature X increases retention because users who enabled it retained 20\% better than users who did not. Identify the flaws in this inference and design a correct analysis.
(2) Given three years of quarterly sales data with a strong seasonal pattern and a one-off promotion in one quarter, determine whether the promotion had a real effect and explain the reasoning.
(3) A city claims its new bus lane reduced average commute times by 12\%. Evaluate whether the claim is supported and identify the main confounds.

\paragraph{Strategic planning.}
(1) Develop a 12-month market-entry plan for a small B2B software company expanding into the Japanese market.
(2) Plan how a university library should reallocate its budget over the next three years as print circulation declines and digital demand grows.
(3) Design a competitive strategy for a mid-size grocery retailer responding to a new low-cost discounter entering its region.

\paragraph{Information synthesis.}
(1) Synthesize the findings of twenty mixed-quality studies on remote-work productivity into a one-page executive briefing with defensible conclusions.
(2) Integrate customer interviews, support tickets, and usage analytics into a single coherent explanation of why users churn from a subscription product.
(3) Three commissioned market-research reports reach conflicting conclusions. Reconcile them into one set of conclusions a board can act on.

\paragraph{Problem solving.}
(1) A distributed job queue intermittently drops tasks under high load. Diagnose the likely causes and propose fixes.
(2) A hospital's patient-discharge process takes twice as long as at peer hospitals. Find the root causes and redesign the process.
(3) A mobile app's crash rate doubled after a minor release with no obvious culprit in the changelog. Work out how to isolate and fix the cause.

\bibliography{references}

\end{document}